# DIFFERENCE–HUFFMAN CODING OF MULTIDIMENSIONAL DATABASES

István Szépkúti

ING Service Centre Budapest Ltd.
H-1068 Budapest, Dózsa György út 84/b, Hungary
e-mail: szepkuti@inf.u-szeged.hu

**Abstract**

A new compression method called difference–Huffman coding (DHC) is introduced in this paper. It is verified empirically that DHC results in a smaller multidimensional physical representation than those for other previously published techniques (single count header compression, logical position compression, base–offset compression and difference sequence compression).

The article examines how caching influences the expected retrieval time of the multidimensional and table representations of relations. A model is proposed for this, which is then verified with empirical data. Conclusions are drawn, based on the model and the experiment, about when one physical representation outperforms another in terms of retrieval time. Over the tested range of available memory, the performance for the multidimensional representation was always much quicker than for the table representation.

*Keywords:* compression, multidimensional database, On-line Analytical Processing, OLAP.

# 1 Introduction

## 1.1 Motivation

Why should we seek to compress multidimensional databases? This is the question we intend to answer in this subsection.

The total number of cells in a multidimensional database equals the product of the number of dimension values $= \prod_{i=1}^{n} |D_i|$, where $D_i$ is the $i^{\text{th}}$ dimension ($i = 1, 2, \ldots, n$) and $n$ is the number of dimensions. This value can be very large, which may make it impractical or even impossible to store all of these cells. The multidimensional databases are usually quite sparse. So we can decrease their size if we get rid of the empty cells. This is the basic idea behind the compression techniques described in Sections 2 and 3.

Now, let us mention why we need increasingly better compression techniques. If a compression technique $A$ can achieve a lower compression ratio than an alternative technique $B$, then more data can be stored into one disk block with $A$ than with $B$. This will result in fewer disk I/O operations when the compressed data are read or written. Firstly, this can be a significant benefit (performance gain) if we replace $B$ with $A$. Secondly, $A$ may be more CPU-intensive than $B$, which is a cost. The balance of the benefits and costs will probably decide which method should be preferred to the other. Thus the overall goal is to find techniques which can produce more benefits than costs.

## 1.2   Results

The results of this paper can be summarized as follows:

- A new compression technique called difference–Huffman coding is presented here.

- It is demonstrated, using experiments on benchmark databases (TPC-D and APB-1), that DHC outperforms other multidimensional compression methods like single count header compression, logical position compression, base–offset compression and difference sequence compression (DSC).

- Just like DSC, DHC is generally able to create smaller databases than the corresponding table representation version compressed with different compression programs. There are only two exceptions – bzip2 and WinRAR – which are better for the APB-1 benchmark database.

- A model is proposed to analyze the caching effects of the alternative physical representations. The model is verified by a number of experiments.

- The experiments also demonstrate that DHC is considerably faster than the table representation when the same amount of memory is available (for pre-loading some parts of the physical representation and for caching).

## 1.3   Related Work

In [17], several related articles have already been mentioned: [2, 4, 5, 7, 9, 12, 13, 14, 16, 18, 19, 21, 24]. Hence we will mention here only those that directly lead to the elaboration of difference–Huffman coding.

The single count header compression (SCHC) was introduced in [2]. A variation of SCHC was described in [16]. In addition to this variation, the paper introduced two new compression techniques: logical position compression (LPC) and base–offset compression (BOC). With LPC, the size of the header can be decreased by 50% when the size of SCHC header is maximal. BOC is able to decrease the header still further. In [17], another compression technique called difference sequence compression was introduced which is able to decrease the header still further in some circumstances. Table 2 and Table 3 are also from [16, 17], apart from those lines of the tables that show additional data on DHC. The single count header compression, logical position compression, base–offset compression and difference sequence compression are described in more details in Section 2.

In the literature, several papers deal with compressed databases: For further details the reader may wish to consult [1, 8, 10, 22, 23].

The paper of Westmann and al. [22] lists more related works in this field. In addition, this article discusses how compression can be integrated into a relational database system. It does not concern itself with the multidimensional physical representation, which is the main focus of the paper. Their key result is that compression can significantly improve the response time of queries if very *light-weight* compression techniques are used. Their benchmark results demonstrate that compression indeed offers high performance gains (up to 50%) for I/O-intensive queries and moderate gains for CPU-intensive queries. Compression can, however, also increase the running time of certain update operations. In this paper we will analyze the retrieval (or point query) operation only, as a lot of On-line Analytical Processing (OLAP) applications handle the data in a *read only* or *read mostly* way. The database is updated outside working hours in batch. Despite this difference, we also encountered performance degradation owing to compression when the entire physical representation was cached into the memory. In this case, in one of the benchmark databases (TPC-D), the multidimensional representation became slower than the table one because of the CPU-intensive Huffman decoding.

Chen et al. [1] propose a Hierarchical Dictionary Encoding for string-valued attributes. The article discusses query optimization issues for compressed databases. Both of these topics are beyond the scope of our paper.

In the article of O'Connell et al. [10], compressing of the data itself is analyzed in a database built on a triple store. It is found that, for some applications, gains in performance of over 50% are attainable, and in OLTP-like[1] situations, there are also gains to be made. This paper deals only with OLAP. We remove the empty cells from the multidimensional array, but do not compress the data itself.

Wu et al. [23] present the theoretical analysis of difference coding for sets and relational tables. The theoretical results were verified with simulations outlined in that paper. Here we combine difference coding with Huffman coding, which results in additional improvements of the compression ratio in the tested benchmark databases.

The normalization of a data cube is the process of choosing an ordering for the attribute (or dimension) values, and the chosen ordering will affect the physical storage of the cube's data. This is the principal theme in the paper by Kaser et al. [8]. The data cube normalization is outside the scope of our article.

The JPEG compression of still images is a wide-spread practice nowadays. The coding process consists of six steps: (1) Block preparation, (2) Discrete cosine transformation, (3) Quantization, (4) Differential quantization, (5) Run-length coding, (6) Statistical coding of the output. For a detailed description of this see [18], for instance. Actually, the last (sixth) step is the Huffman coding of the result produced by the previous one. We also apply two different compressions (difference and Huffman codings) one after the other in order to better the compression ratio. However, we here compress the sequence of logical positions. Hence our method is lossless, unlike JPEG, which can be lossy as well for example in steps (1) and (3).

When we analyze algorithms, which operate on data on the secondary stor-age, we usually examine how many disk I/O operations are required by the algorithm. This is because we follow the *dominance of the I/O cost* rule [3]. During our analysis of the caching effects, we approached the question differently. Instead of counting the number of disk I/O operations, we introduced two different constants ($D_m$ and $D_t$) and determined them with experiments. The tests showed that $D_m \ll D_t$, which means of course that more disk I/O operations are needed to retrieve one row from the table representation than one cell from the multidimensional representation when there is no caching. However, for our model, it was not necessary to know the exact number of I/O operations for the alternative physical representations.

---

[1]OLTP stands for On-line Transaction Processing.

## 1.4 Organization

The rest of the paper is organized as follows. Section 2 describes four previously published compression techniques: single count header compression, logical position compression, base–offset compression and difference sequence compression. Section 3 introduces an improved method, that of difference–Huffman coding. The effect of caching alternative physical representations is analyzed in Section 4. The theoretical results are then tested in experiments outlined in Section 5. Section 6 rounds off the discussion with some conclusions and suggestions for future study. Lastly, for completeness, we have the Acknowledgements and an appendix section followed by a list of references.

## 2  Compression Techniques

Throughout this paper we employ the expressions "multidimensional representation" and "table representation", which are defined as follows.

**Definition.** Suppose we wish to represent relation $R$ physically. The multidimensional (physical) representation of $R$ is as follows:

- A compressed array, which only stores the nonempty cells, one nonempty cell corresponding to one element of $R$;

- The header, which is needed for the logical-to-physical position transformation;

- One array per dimension in order to store the dimension values.

The table (physical) representation consists of the following:

- A table, which stores every element of relation $R$;

- A B-tree index to speed up the access to given rows of the table when the entire primary key is given.

□

The difference–Huffman coding is closely related to single count header compression, logical position compression, base–offset compression and difference sequence compression. The latter techniques are described in the remaining part of this section.

*Single count header compression.* By transforming the multidimensional array into a one-dimensional array, we obtain a sequence of empty and nonempty cells:

$$(E^*F^*)^* \tag{1}$$

In the above regular expression, $E$ is an empty cell and $F$ is a nonempty one. The single count header compression stores only the nonempty cells and the cumulated run lengths of empty cells and nonempty cells. In [15], we made use of a variant of SCHC. The difference between the two methods is that the original method accumulates the number of empty cells and number of nonempty cells separately. These accumulated values are placed in a single alternating sequence. The sum of two consecutive values corresponds to a logical position. (The logical position is the position of the cell in the multidimensional array before compression. The physical position is the position of the cell in the compressed array.) In [15], instead of storing a sequence of values, we chose to store pairs of a logical position and the number of empty cells up to this logical position: $(L_j, V_j)$. Just one pair is stored per $E^*F^*$ run and $L_j$ points to the last element of the corresponding run. From here on when we mention SCHC we only refer to the variant of this compression scheme defined in [15].

**Definition.** The array storing the $(L_j, V_j)$ pairs of logical positions and number of empty cells will be called the SCHC header. □

The following three compression techniques better SCHC when the SCHC header is maximal.

*Logical position compression.* The size of the SCHC header depends on the number of $E^*F^*$ runs. In the worst case there are $N = |R|$ runs, where $R$ is the relation, which is represented multidimensionally using SCHC. Then the size of the SCHC header is $2N\iota$. (We assume that $L_j$ and $V_j$ are of the same data type and each of them occupies $\iota$ bytes of memory.) But then it is better to build another type of header. Instead of storing the $(L_j, V_j)$ pairs, it is more convenient to store just the $L_j$ sequence of each cell (that is, not merely the $L_j$ sequence of runs).

**Definition.** The compression method, which just uses the sequence of logical positions, will be called logical position compression (LPC). The $L_j$ sequence used in logical position compression will be called the LPC header. □

The number of $E^*F^*$ runs lies between 1 and $N = |R|$. Let $\nu$ denote the number of runs. Because the size of $L_j$ and $V_j$ is the same, the header is smaller with a logical position compression if $\frac{N}{2} < \nu$. Otherwise, if $\frac{N}{2} \geq \nu$ a logical position compression does not result in a smaller header than the single count header compression. The header with a logical position compression is half that of the SCHC header in the worst case, that is when $\nu = N$.

*Base–offset compression.* In order to store the entire $L_j$ sequence, we may need a huge (say 8-byte) integer number. However, the sequence is strictly increasing:

$$L_0 < L_1 < \cdots < L_{N-1}. \tag{2}$$

Here, $N$ denotes the number of elements in the $L_j$ sequence. The difference sequence, $\Delta L_j$, contains significantly smaller values. Based on this observation, we may compress the header still further.

Suppose that we need $\iota$ bytes to store one element of the $L_j$ sequence. In addition, there exists a natural number $l$ such that for all $k = 0, 1, 2, \ldots$ the

$$L_{(k+1)l-1} - L_{kl} \tag{3}$$

values may be stored in $\theta$ bytes and $\theta < \iota$. In this case we can store two sequences instead of $L_j$, as it can be seen from the definition below.

**Definition.** For convenience, let

$$B_k = L_{kl}, \tag{4}$$

$$O_j = L_j - B_{\lfloor \frac{j}{l} \rfloor}, \tag{5}$$

where $k = 0, \ldots, \lfloor \frac{N-1}{l} \rfloor$ and $j = 0, \ldots, N-1$. Sequence $B_k$ will be called the base sequence, and sequence $O_j$ will be called the offset sequence. The compression method based on these two sequences will be named base–offset compression (BOC). The base and the offset sequences together will be called the BOC header. □

More details about these three compression techniques can be found in [2, 15, 16, 17].

*Difference sequence compression.* We will now discuss DSC in more detail as it forms the basis of DHC.

The main idea behind DSC is that more flexibility is possible when an absolute address is stored, namely – where necessary –, that is the relative address (offset) might be too large to store on given $s$ bits.

The sequence of logical positions is strictly increasing:

$$L_0 < L_1 < \cdots < L_{N-1}.$$

In addition, the difference sequence $\Delta L_j$ contains smaller values than the original $L_j$ sequence. This property was utilized by base–offset compression and will be used by the difference sequence compression as well.

During the design of the data structures and the search algorithm, the following principles were used:

- We compress the header such that the decompression is quick.

- It is not necessary to decompress the entire header.

- Searching can be done during decompression, and the decompression stops immediately when the header element is found or when it is demonstrated that the header element cannot be found (that is, when the corresponding cell is empty).

**Definition.** Let us introduce the following notation.
$N$ is the number of elements in the sequence of logical positions ($N > 0$);
$L_j$ is the sequence of logical positions ($0 \leq j \leq N - 1$);
$\Delta L_0 = L_0$;
$\Delta L_j = L_j - L_{j-1}, j = 1, 2, \ldots, N - 1$;
The $D_i$ sequence ($D_i \in \{0, 1, \ldots, \overline{D}\}, i = 0, 1, \ldots, N - 1$) is defined as follows:

$$D_i = \begin{cases} \Delta L_i, & \text{if } \Delta L_i \leq \overline{D} \text{ and } i > 0; \\ 0, & \text{otherwise}; \end{cases} \tag{6}$$

where $\overline{D} = 2^s - 1$ and $s$ is the size of a $D_i$ sequence element in bits.

The $J_k$ sequence will be defined recursively in the following way:

$$J_k = \begin{cases} L_0, & \text{if } k = 0; \\ L_j, & \text{otherwise where } j = min\{i \mid \Delta L_i > \overline{D} \text{ and } L_i > J_{k-1}\}. \end{cases} \quad (7)$$

Here the $D_i$ sequence is called the overflow difference sequence. There is an obvious distinction between $\Delta L_i$ and $D_i$, but the latter will also be called the difference sequence, if it is not too disturbing. As for $J_k$ it is called the jump sequence. The compression method which makes use of the $D_i$ and $J_k$ sequences will be called difference sequence compression (DSC). The $D_i$ and $J_k$ sequences together will be called the DSC header. □

Notice here that $\Delta L_i$ and $D_i$ are basically the same sequence. The only difference is that some elements of the original difference sequence $\Delta L_i$ are replaced with zeros, if and only if they cannot be stored in $s$ bits.

The difference sequence will also be called the relative logical position sequence, and we shall call the jump sequence the absolute logical position sequence.

From the definitions of $D_i$ and $J_k$, one can see clearly that, for every zero element of the $D_i$ sequence, there is exactly one corresponding element in the $J_k$ sequence. For example, let us assume that $D_0 = D_3 = D_5 = 0$, and $D_1, D_2, D_4, D_6, D_7, D_8 > 0$. Then the above mentioned correspondence is shown in the following table:

| $D_0$ | $D_1$ | $D_2$ | $D_3$ | $D_4$ | $D_5$ | $D_6$ | $D_7$ | $D_8$ | ... |
|---|---|---|---|---|---|---|---|---|---|
| $J_0$ | | | $J_1$ | | $J_2$ | | | | ... |

From the above definition, the recursive formula below follows for $L_j$.

$$L_j = \begin{cases} L_{j-1} + D_j, & \text{if } D_j > 0; \\ J_k, & \text{otherwise where } k = min\{i \mid J_i > L_{j-1}\}. \end{cases} \quad (8)$$

In other words, every element of the $L_j$ sequence can be calculated by adding zero or more consecutive elements of the $D_i$ sequence to the proper jump sequence element. For instance, in the above example

$L_0 = J_0$;
$L_1 = J_0 + D_1$;
$L_2 = J_0 + D_1 + D_2$;
$L_3 = J_1$;
$L_4 = J_1 + D_4$;
and so on.

Now the number of elements in the offset array and the difference array is just the same, but are there fewer jumps than base array elements? The answer to this question is that there are no more jumps than base array elements when the size of one offset array element ($\theta$) is less than or equal to the size of one difference array element ($\zeta$).

**Theorem 1.** There are never more jumps than base array elements if $\theta \leq \zeta$.
The proof of this is given in [17].

**Corollary.** The multidimensional representation with DSC does not result in a bigger database size than with BOC if $\theta = \zeta$.

In order to find a given $L$ quickly (using the DSC header) in the $L_j$ sequence when the corresponding cell is not empty, we need an $A_k$ sequence of pointers which is defined as follows.

**Definition.** For every $k$, $A_k = j$, if and only if $J_k = L_j$. We will refer to the $A_k$ sequence as the accelerator sequence.   □

**Corollary.** Suppose $J_k$ is an element of the jump sequence. Then the corresponding difference sequence element is $D_{A_k}$, which equals zero by definition. Thus the accelerator sequence can be employed to find the corresponding difference sequence element of a jump quite quickly.

In order to save space we can modify the above definition of $A_k$ and store only $A_0, A_n, A_{2n}, \ldots$, that is just every $n^{\text{th}}$ element of the original accelerator sequence.

In this case, in the searching algorithm, we have to expect zero difference sequence elements as well. When a zero comes, we will take the next element of the jump sequence. However, at the beginning of the algorithm it is quite sufficient to find $L$ with a binary search among the elements $J_0, J_n, J_{2n}, \ldots$ because the accelerator sequence only contains pointers for these jumps.

The accelerator sequence is a useful method for speeding up the retrieval (point query) operation for the following reasons:

- It is not necessary to store the accelerator sequence on the hard disk since it can be easily populated based on the difference sequence in one pass. This is needed only once after the difference array is loaded from the hard disk into the memory.

- In practice the sequence does not increase the memory requirements significantly, as was shown in [17].

A detailed analysis of DSC and the search algorithm are in [17] as well.

## 3   Difference – Huffman Coding

The key idea in difference – Huffman coding is that we can compress the difference sequence further if we replace it with its corresponding Huffman code.

**Definition.** The compression method, which uses the jump sequence ($J_k$) and the Huffman code of the difference sequence ($D_i$), will be labelled difference – Huffman coding (DHC). The $J_k$ sequence and the Huffman code of the $D_i$ sequence together will be called the DHC header.   □

The difference sequence usually contains a lot of *zeros*. Moreover, it contains as many *ones* too if there are numerous consecutive elements in the $L_j$ sequence of logical positions. By definition, the elements of the difference sequence are smaller than those of the logical position sequence. The elements of $D_j$ will recur with greater or less frequency. Hence it seems reasonable to code the frequent elements with fewer bits, and the less frequent ones with more. To do this, the optimal prefix code can be determined by the well-known Huffman algorithm [6].

In the case of DSC, the accelerator sequence stores those indices that can be used to access the difference sequence. This is different in DHC, as only the Huffman code of the difference sequence can be found in the memory. In addition to the accelerator sequence (or array), two more arrays are needed. These are:

- One array, which stores the pointers to given *bytes* in the Huffman code of the difference sequence;

- Another one to store the *bit* position, where the given element of the difference sequence ends within the aforementioned byte.[2]

Figure 1: The Huffman code of the difference sequence

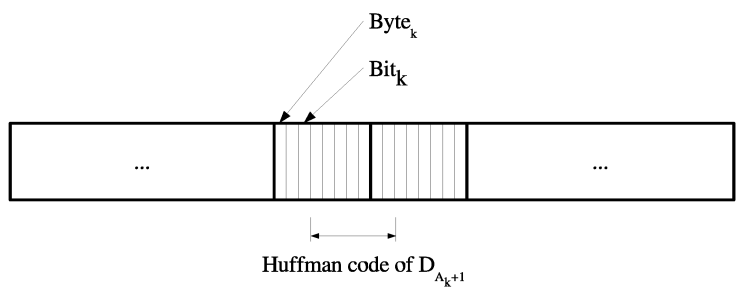

There is a correspondence between the accelerator and difference sequence elements of DSC. For instance, see Table 1.

Table 1:

| $A_0$ |  |  | $A_1$ |  | $A_k$ |  |  | $A_{k+1}$ | ... |
|---|---|---|---|---|---|---|---|---|---|
| $D_0$ | $D_1$ | $D_2$ | $D_3$ | ... | $D_{A_k}$ | $D_{A_k+1}$ | ... | $D_{A_{k+1}}$ | ... |

In this example, $D_0 = D_3 = D_{A_k} = D_{A_{k+1}} = 0$, by definition.

The situation is different with DHC, as can be seen in Figure 1. The figure shows the Huffman code of the difference sequence. $Byte_k$ points to the byte position where the Huffman code of $D_{A_k}$ ends. Similarly, $Bit_k$ points to the bit within the byte where the previous code finishes. Right after this bit, the Huffman code of $D_{A_k+1}$ begins. Calling the Huffman decoder from here, the value of $D_{A_k+1}$ can be determined.

---

[2]Similarly, we could store the bit position where the Huffman code of the difference sequence element in question starts. We chose to store the position of the end bit, because it was already available, when the three arrays were populated with values. No additional calculations were required; so the populating of the arrays became simpler. This choice also helps when the next element of the difference sequence is needed, as it starts right after the current one, from the next bit position.

**Remark.** Firstly, in DSC, $A_k$ points to the difference sequence element ($D_{A_k}$), which corresponds to $J_k$. Secondly, using the byte and bit positions, we can decode $D_{A_k+1}$ instead of $D_{A_k}$ in DHC. This change does not cause any problem as $D_{A_k} = 0$ by definition.

Applying DSC, we can find a cell in the multidimensional physical representation with the following procedure.

- Using the DSC header, find the difference sequence element ($D_j$) for which the following equation holds.

$$L = J_k + D_{A_k} + D_{A_k+1} + D_{A_k+2} + \cdots + D_{j-1} + D_j, \qquad (9)$$

  where $J_k$ is a jump and $D_{A_k+1}, D_{A_k+2}, \ldots, D_{j-1}$ and $D_j$ are consecutive positive difference sequence elements. $L$ is the logical position of the cell we are looking for.

- If such a $D_j$ cannot be found, then the cell is left empty.

- Otherwise $j$ is the physical position corresponding to the logical position $L$, and the content of the cell can be found at this physical location in the compressed multidimensional array.

We cannot do exactly the same if the multidimensional representation is compressed with DHC. The reason for this is that the Huffman code of the difference sequence is used instead of the original difference sequence. In this case, the search algorithm works like this:

- Using the jump sequence, find the largest jump $J_k$, for which the following inequality is true:

$$J_k \leqq L, \qquad (10)$$

  where $L$ is the logical position of the cell we are seeking. If such a $J_k$ cannot be found, the cell is left empty.

- Initialize the Huffman decoder with $Byte_k$ and $Bit_k$. While the

$$J_k + D_{A_k+1} + D_{A_k+2} + \cdots + D_{A_k+\ell} < L \qquad (11)$$

  inequality holds and before we reach the end of the difference sequence, decode the difference sequence elements ($D_{A_k+1}, D_{A_k+2}, \ldots$) one by one.

- If the end of the difference sequence has been reached or

$$J_k + D_{A_k+1} + D_{A_k+2} + \cdots + D_{A_k+\ell} > L, \qquad (12)$$

  then the logical position $L$ cannot be found and the cell is left empty.

- Otherwise, if

$$J_k + D_{A_k+1} + D_{A_k+2} + \cdots + D_{A_k+\ell} = L, \qquad (13)$$

  then we have found the logical position $L$ after adding $\ell$ consecutive positive difference sequence elements to $J_k$ ($\ell = 0, 1, \ldots$). Hence the physical position of the cell in the compressed multidimensional array is $A_k + \ell$.

Similar to DSC, it is not necessary to store every element of the $A_k$, $Byte_k$ and $Bit_k$ arrays. To save space we can modify the above search algorithm slightly and store only $A_0, A_n, A_{2n}, \ldots, Byte_0, Byte_n, Byte_{2n}, \ldots$ and $Bit_0, Bit_n, Bit_{2n}, \ldots$, that is just every $n^{\text{th}}$ element of the original arrays. This is what happened in the experiments, where $n$ was equal to 16.

# 4 Caching

In this section we shall examine how the caching affects the speed of retrieval in the different physical database representations. For the analysis, a model will be proposed. Then we will give sufficient and necessary conditions for when the expected retrieval time is smaller in one representation than in the other.

The caching can speed up the operation of a database management system significantly if the same block is requested while it is still in the memory. In order to show how the caching modifies the results of this paper, let us introduce the following notations.

**Definition.**

$$
\begin{aligned}
M &= \text{The retrieval time, if the information is in the memory.} \\
D &= \text{The retrieval time, if the disk also has to be accessed.} \\
p &= \text{The probability of having everything needed in the memory.} \\
q &= 1 - p \\
\xi &= \text{How long it takes to retrieve the requested information.}
\end{aligned}
$$

$\square$

In our model we shall consider $M$ and $D$ constants. Obviously, $\xi$ is a random variable. Its expected value can be calculated as follows:

$$\mathbb{E}(\xi) = pM + qD \tag{14}$$

Notice that $D$ does not tell us how many blocks have to be read from the disk. This also means that the value of $D$ will be different for the table and the multidimensional representations. The reason for this is that, in general, at most one block has to be read with the multidimensional representation. Exactly one reading is necessary if nothing is cached, because only the compressed multidimensional array is kept on the disk. Everything else (the header, the dimension values, and so forth) is loaded into the memory in advance. With the table representation, more block readings may be needed because we also have to traverse through the B-tree first, and then we have to retrieve the necessary row from the table.

$M$ is also different for the two alternative physical representations. This is so because two different algorithms are used to retrieve the same information from two different physical representations.

Hence, for the above argument, we are going to introduce four constants.

**Definition.**

$$M_m = \text{The value of } M \text{ for the multidimensional representation.}$$
$$M_t = \text{The value of } M \text{ for the table representation.}$$
$$D_m = \text{The value of } D \text{ for the multidimensional representation.}$$
$$D_t = \text{The value of } D \text{ for the table representation.}$$

□

If we sample the cells/rows with uniform probability[3], we can then estimate the probabilities as follows:

$$p = \frac{\text{The size of the cached blocks of the physical representation}}{\text{The total size of the physical representation}} \quad (15)$$

$$q = 1 - p \quad (16)$$

By the "total size" we mean that part of the physical representation which can be found on the disk at the beginning. In the multidimensional representation, it is the compressed multidimensional array, whereas in the table representation, we can put the entire size of the physical representation into the denominator of $p$. The cached blocks are those that had been originally on the disk, but were moved into the memory later. In other words, the size of the cached blocks (numerator) is always smaller than or equal to the total size (denominator).

The experiments shows that the alternative physical representations differ from each other in size. That is why it seems reasonable to introduce four different probabilities in the following manner.

**Definition.**

$$p_m = \text{The value of } p \text{ for the multidimensional representation.}$$
$$p_t = \text{The value of } p \text{ for the table representation.}$$
$$q_m = 1 - p_m$$
$$q_t = 1 - p_t$$

□

When does the inequality below hold? This is an important questions.

$$\mathbb{E}(\xi_m) < \mathbb{E}(\xi_t) \quad (17)$$

Here $\xi_m$ and $\xi_t$ are random variables that are the retrieval times in the multidimensional and table representations, respectively.

In our model, $\mathbb{E}(\xi_i) = p_i M_i + q_i D_i$ ($i \in \{m, t\}$). Thus the question can be rephrased as follows.

$$p_m M_m + q_m D_m < p_t M_t + q_t D_t \quad (18)$$

The value of the $M_m$, $D_m$, $M_t$ and $D_t$ constants was measured by carrying out some experiments. (See the following section.) Two different results were obtained. For one benchmark database (TPC-D), the following was found.

$$M_t < M_m \ll D_m \ll D_t \quad (19)$$

---

[3]Here and in the remainder of the paper we shall make the same assumption that every cell/row is sampled with the same probability.

The other database (APB-1) gave a slightly different result.

$$M_m \ll M_t \ll D_m \ll D_t \tag{20}$$

The second pair of inequalities ($M_m \ll D_m$ and $M_t \ll D_m$) can be accounted for by the fact that disk operations are slower than memory operations with orders of magnitude. The third one ($D_m \ll D_t$) is because we have to retrieve more blocks from the table representation than from the multidimensional to obtain the same information.

Note here that $\mathbb{E}(\xi_i)$ is the convex linear combination of $M_i$ and $D_i$ ($p_i, q_i \in [0,1]$ and $i \in \{m,t\}$). In other words, $\mathbb{E}(\xi_i)$ can take any value from the closed interval $[M_i, D_i]$.

The following provides a sufficient condition for $\mathbb{E}(\xi_m) < \mathbb{E}(\xi_t)$.

$$D_m < p_t M_t + q_t D_t \tag{21}$$

From this, with equivalent transformations, we obtain the inequality constraint:

$$D_m < p_t M_t + (1-p_t) D_t \tag{22}$$

$$p_t < \frac{D_t - D_m}{D_t - M_t} \tag{23}$$

The value for $\frac{D_t - D_m}{D_t - M_t}$ was found to be 63.2% (TPC-D) and 66.3% (APB-1) in the experiments. This means that, based on the experimental results, the expected value of the retrieval time was smaller in the multidimensional representation than in the table representation when less than 63.2% of the latter one was cached. This was true regardless of the fact of whether the multidimensional representation was cached or not.

Now we are going to differentiate two cases based on the value of $M_m$ and $M_t$.

*Case 1:* $M_t < M_m$. This was true for the TPC-D benchmark database. (Here the difference sequence consisted of 16-bit unsigned integers, which resulted in a slightly more complicated decoding, as the applied Huffman decoder returns 8 bits at a time. This may be the reason why $M_m$ became larger than $M_t$.) In this case, we can give a sufficient condition for $\mathbb{E}(\xi_m) > \mathbb{E}(\xi_t)$, as the equivalent transformations below show.

$$p_t M_t + q_t D_t < M_m \tag{24}$$
$$p_t M_t + (1-p_t) D_t < M_m \tag{25}$$
$$\frac{D_t - M_m}{D_t - M_t} < p_t \tag{26}$$

For $\frac{D_t - M_m}{D_t - M_t}$ we obtained a value of 99.9%. This means that the expected retrieval time was smaller in the *table* representation when more than 99.9% of it was cached. This was true even when the whole multidimensional representation was in the memory.

*Case 2:* $M_m \ll M_t$. This inequality held true for the APB-1 benchmark database. Here we can give another sufficient condition for $\mathbb{E}(\xi_m) < \mathbb{E}(\xi_t)$.

$$p_m M_m + q_m D_m < M_t \tag{27}$$
$$p_m M_m + (1-p_m) D_m < M_t \tag{28}$$
$$\frac{D_m - M_t}{D_m - M_m} < p_m \tag{29}$$

The left hand side of the last inequality was equal to 98.3% for the APB-1 benchmark database. In other words when more than 98.3% of the multidimensional representation was cached, it then resulted in a faster operation on average than the table representation regardless of the caching level of the latter.

Finally, let us give a necessary and sufficient condition for $\mathbb{E}(\xi_m) < \mathbb{E}(\xi_t)$. First, let us consider the following equivalent transformations (making the natural assumption that $D_t > M_t$).

$$\mathbb{E}(\xi_m) < \mathbb{E}(\xi_t) \tag{30}$$
$$p_m M_m + q_m D_m < p_t M_t + q_t D_t \tag{31}$$
$$p_m M_m + (1 - p_m) D_m < p_t M_t + (1 - p_t) D_t \tag{32}$$
$$p_t < \frac{D_m - M_m}{D_t - M_t} p_m + \frac{D_t - D_m}{D_t - M_t} \tag{33}$$

The last inequality was the following for the two tested databases, TPC-D and APB-1, respectively:

$$p_t < 0.368 p_m + 0.632 \tag{34}$$
$$p_t < 0.343 p_m + 0.663 \tag{35}$$

**Theorem 2.** Suppose that $D_t > M_t$. Then the expected retrieval time is smaller in the case of the multidimensional physical representation than in the table physical representation if and only if

$$p_t < \frac{D_m - M_m}{D_t - M_t} p_m + \frac{D_t - D_m}{D_t - M_t}. \tag{36}$$

The truth of the theorem is a direct consequence of equations (30)–(33).

We conclude this section by summarizing our findings:

- The caching of the alternative physical representations modify the results significantly.

- If (nearly) the entire physical representation is cached into the memory, then the complexity of the algorithm will determine the speed of retrieval. The less CPU-intensive algorithm will probably result in a faster operation.

- In the tested cases, the expected retrieval time was smaller with multidimensional physical representation when less than 63.2% of the table representation was cached. This was true regardless of the caching level of the multidimensional representation.

## 5 Experiments

We carried out experiments in order to measure the sizes of the different physical representations and the constants in the previous section. We also examined how the size of the cache influenced the speed of retrieval. The hardware and software components we used for our experiments are listed in the appendix section.

In the experiments we made use of two benchmark databases: TPC-D [20] and APB-1 [11]. One relation was derived per benchmark database in exactly the same way as that described in [16]. Then these relations were represented physically with a multidimensional representation and table representation.

When we compare the DHC of the multidimensional representation of relation $R$ to compressions of the table representation of relation $R$ we get an interesting result. (Here $R$ is a relation derived from one of the benchmark databases: TPC-D or APB-1.) Both Table 2 and Table 3 show that DHC results in a smaller multidimensional representation than difference sequence compression. With the TPC-D benchmark database, the multidimensional representation with BOC and DSC turned out to be already smaller than all those used for alternative compression techniques of the table representation (see [16]).

In the APB-1 benchmark database, BOC was less successful. It produced a slightly larger database than the compressions of the table representation. However, with the exception of bzip2 and WinRAR, DSC outperformed the other compressors. Obviously this observation is true for DHC as well.

In both benchmark databases, DHC produced the smallest multidimensional physical representation.

Table 2: TPC-D benchmark database

| Compression | Size in bytes | Percentage |
|---|---|---|
| **Table representation** | | |
| Uncompressed | 279,636,324 | 100.0% |
| ARJ | 92,429,088 | 33.1% |
| gzip | 90,521,974 | 32.4% |
| WinZip | 90,262,164 | 32.3% |
| PKZIP | 90,155,633 | 32.2% |
| jar | 90,151,623 | 32.2% |
| bzip2 | 86,615,993 | 31.0% |
| WinRAR | 81,886,285 | 29.3% |
| **Multidimensional representation on the disk** | | |
| Single count header compression | 145,256,792 | 51.9% |
| Base–offset compression | 74,001,692 | 26.5% |
| Difference sequence compression | 67,925,100 | 24.3% |
| Difference–Huffman coding | 66,556,350 | 23.8% |
| **Multidimensional representation in the memory** | | |
| Difference–Huffman coding | 67,014,312 | 24.0% |

As we explained earlier in this paper, the size of the multidimensional representation with DHC is different on the disk and in the memory. This is because of the existence of the $A_k$, $Byte_k$ and $Bit_k$ arrays. The last lines of Table 2 and Table 3 show the memory occupancy of DHC. We can arrange it such that these three arrays do not increase the memory requirements of the multidimensional physical representation significantly.

In the rest of this section, we shall deal only with DHC. Its performance will be compared to the performance of the uncompressed table representation.

Table 3: APB-1 benchmark database

| Compression | Size in bytes | Percentage |
|---|---|---|
| **Table representation** | | |
| Uncompressed | 1,295,228,960 | 100.0% |
| jar | 124,462,168 | 9.6% |
| gzip | 124,279,283 | 9.6% |
| WinZip | 118,425,945 | 9.1% |
| PKZIP | 117,571,688 | 9.1% |
| ARJ | 115,085,660 | 8.9% |
| bzip2 | 99,575,906 | 7.7% |
| WinRAR | 98,489,368 | 7.6% |
| **Multidimensional representation on the disk** | | |
| Base–offset compression | 125,572,184 | 9.7% |
| Difference sequence compression | 113,867,897 | 8.8% |
| Single count header compression | 104,959,936 | 8.1% |
| Difference–Huffman coding | 103,072,522 | 8.0% |
| **Multidimensional representation in the memory** | | |
| Difference–Huffman coding | 103,369,039 | 8.0% |

In order to determine the constant values of the previous section, another experiment was performed. A random sample was taken with replacement from relation $R$ with uniform distribution. The sample size was 1000. Afterwards the sample elements were retrieved from the multidimensional representation and then from the table representation. The elapsed time was measured to calculate the average retrieval time per sample element. Then the same sample elements were retrieved again from the two physical representations. Before the first round, nothing was cached. So the results help us to determine the constants $D_m$ and $D_t$. Before the second round, every element of the sample was cached in both physical representations. So the times measured in the second round correspond to the values of the constants $M_m$ and $M_t$. The results of the experiment can be seen in Table 4.

Table 4: Constants

| Symbol | TPC-D (ms) | APB-1 (ms) |
|---|---|---|
| $M_m$ | 0.031 | 0.012 |
| $M_t$ | 0.021 | 0.128 |
| $D_m$ | 6.169 | 6.778 |
| $D_t$ | 16.724 | 19.841 |

In the next experiment, we examined how the size of memory available for caching influenced the speed of retrieval. But first we should mention what we expect to get based on our model. With the multidimensional representation,

DIFFERENCE–HUFFMAN CODING OF MULTIDIMENSIONAL DATABASES        17

the formula below follows from the model for the expected retrieval time:

$$T_m(x) = M_m p_m(x) + D_m q_m(x) = M_m p_m(x) + D_m(1 - p_m(x)), \qquad (37)$$

where

$$p_m(x) = \min\left\{\frac{x-H}{C}, 1\right\}, \qquad (38)$$

$H$ is the total size of the multidimensional representation part, which is loaded into the memory in advance (the jump array, the Huffman code of the difference sequence, the decoding tree, the dimension values, the $A_k$, $Byte_k$ and $Bit_k$ arrays), $C$ is the size of the compressed multidimensional array and $x$ ($\geqq H$) is the size of the available memory.

In an analogous way for the table representation, we obtain the formula:

$$T_t(x) = M_t p_t(x) + D_t q_t(x) = M_t p_t(x) + D_t(1 - p_t(x)), \qquad (39)$$

where

$$p_t(x) = \min\left\{\frac{x}{S}, 1\right\}, \qquad (40)$$

$S$ is the total size of the table representation and $x$ ($\geqq 0$) is the size of the memory available for caching.

In Figure 2 and Figure 3, $T_m(x)$ is labelled as "Array Est", $T_t(x)$ as "Table Est". The horizontal axis shows the size of the memory in bytes, while the vertical one displays the expected/average retrieval time in milliseconds.

It is not hard to see that the global maximum and minimum values and locations of the functions $T_m(x)$ and $T_t(x)$ are the following:

$$\max\{T_m(x) \mid x \geqq H\} = D_m \quad \text{and} \quad T_m(x) = D_m \quad \text{if and only if} \quad x = H$$

$$\min\{T_m(x) \mid x \geqq H\} = M_m \quad \text{and} \quad T_m(x) = M_m \quad \text{if and only if} \quad x \geqq H + C$$

$$\max\{T_t(x) \mid x \geqq 0\} = D_t \quad \text{and} \quad T_t(x) = D_t \quad \text{if and only if} \quad x = 0$$

$$\min\{T_t(x) \mid x \geqq 0\} = M_t \quad \text{and} \quad T_t(x) = M_t \quad \text{if and only if} \quad x \geqq S$$

In order to verify the model with empirical data, we arranged the following tests. Random samples were taken with replacement. The sample size was set at 300 in TPC-D and 100 in APB-1 in order to stay within the constraints of the physical memory. The average retrieval time was measured as well as the cache size used for each physical representation. In the multidimensional representation, the utilized cache size was corrected by adding $H$ to it, as this representation requires that some parts of it are loaded into the memory in advance. Then the above sampling and measuring procedures were repeated another 99 times. That is, altogether 30,000 elements were retrieved from the TPC-D database and 10,000 from the APB-1. The average retrieval time, as a function of the cache size (or memory) used, can also be seen in Figures 2 and 3. The data relating to the multidimensional physical representation are labelled as "Array", and the data for the table representation as "Table".

Both diagrams suggest that the model fits the empirical data quite well. Only the table representation of ABP-1 deviates slightly from it.

Figure 2: The retrieval time for the TPC-D benchmark database as a function of the memory size available for caching

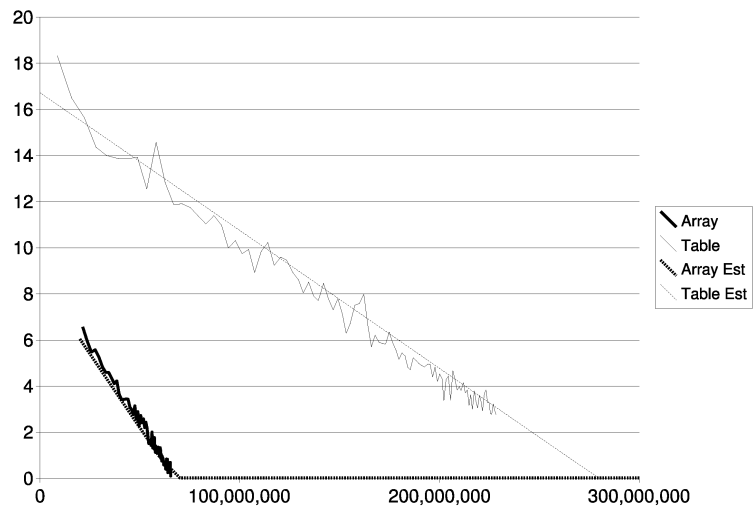

Figure 3: The retrieval time for the APB-1 benchmark database as a function of the memory size available for caching

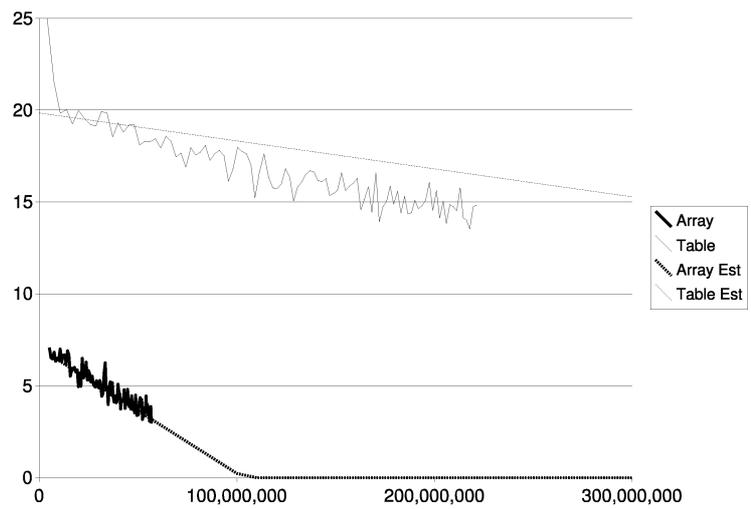

The test results of the first ten passes and the last ten passes can be seen in Table 5 as well. Column A is the sequence number. Columns B–E correspond to TPC-D, while columns F–I are for APB-1. Columns B and F show the memory needed for the multidimensional representation, while columns C and G give the same for the table representation. The retrieval time with the multidimensional representation can be found in columns D and H, and the table representation in columns E and I. The "memory used" values are strictly increasing. This can be attributed to the fact that increasingly larger parts of the physical representations are cached into to the memory.

Looking at Table 5, Figure 3 and Figure 4, it can be seen that the multidimensional representation was always significantly faster over the tested range.

Table 5: Memory used (in $2^{10}$ bytes) and retrieval time (in milliseconds) for the TPC-D and the APB-1 benchmark databases

| A | B | C | D | E | F | G | H | I |
|---|---|---|---|---|---|---|---|---|
| 1 | 20,893 | 8,500 | 6.57 | 18.32 | 4,926 | 3,840 | 7.10 | 24.99 |
| 2 | 23,093 | 15,488 | 5.96 | 16.50 | 5,698 | 7,204 | 6.55 | 21.53 |
| 3 | 25,097 | 21,732 | 5.48 | 15.64 | 6,478 | 10,312 | 6.48 | 19.83 |
| 4 | 27,025 | 27,420 | 5.58 | 14.36 | 7,262 | 13,452 | 6.85 | 20.03 |
| 5 | 28,841 | 32,668 | 5.26 | 14.00 | 8,002 | 16,328 | 6.35 | 19.25 |
| 6 | 30,565 | 37,896 | 4.83 | 13.88 | 8,774 | 19,336 | 6.52 | 19.99 |
| 7 | 32,113 | 42,908 | 4.61 | 13.87 | 9,506 | 22,208 | 6.42 | 19.56 |
| 8 | 33,557 | 47,684 | 4.60 | 13.92 | 10,266 | 25,076 | 7.02 | 19.23 |
| 9 | 34,949 | 52,228 | 4.37 | 12.56 | 10,978 | 27,884 | 6.35 | 19.13 |
| 10 | 36,289 | 56,792 | 4.12 | 14.58 | 11,726 | 30,664 | 6.68 | 19.92 |
| ⋮ | ⋮ | ⋮ | ⋮ | ⋮ | ⋮ | ⋮ | ⋮ | ⋮ |
| 91 | 63,609 | 216,352 | 0.35 | 2.94 | 52,334 | 201,140 | 3.72 | 13.82 |
| 92 | 63,677 | 217,228 | 0.70 | 3.69 | 52,726 | 202,836 | 4.46 | 14.86 |
| 93 | 63,729 | 218,060 | 0.24 | 3.83 | 53,046 | 204,540 | 3.55 | 14.75 |
| 94 | 63,769 | 218,784 | 0.22 | 3.29 | 53,438 | 206,240 | 3.98 | 14.52 |
| 95 | 63,813 | 219,484 | 0.28 | 3.31 | 53,754 | 207,960 | 3.47 | 15.77 |
| 96 | 63,841 | 220,200 | 0.34 | 2.82 | 54,090 | 209,516 | 3.82 | 14.12 |
| 97 | 63,857 | 220,804 | 0.13 | 2.78 | 54,382 | 211,100 | 3.09 | 14.01 |
| 98 | 63,905 | 221,592 | 0.30 | 3.23 | 54,670 | 212,660 | 3.13 | 13.53 |
| 99 | 63,925 | 222,260 | 0.11 | 2.94 | 55,054 | 214,404 | 3.89 | 14.74 |
| 100 | 63,949 | 222,908 | 0.32 | 2.78 | 55,358 | 216,144 | 2.97 | 14.83 |

Summarizing our experimental results, we may say that

- The size of DHC was smaller than that of the other compressed multidimensional representations. This was true even when we included those parts of DHC that were not stored on the disk, but recalculated every time the header was loaded into the memory.

- With suitably designed experiments, we were able to measure the constants of the model proposed in the previous section.

- We tested the model with empirical data.

- Over the tested range of available memory, the multidimensional representation was always much quicker than the table representation in terms of retrieval time.

# 6 Conclusion

In this paper we introduced a new compression method called difference–Huffman coding. In our experiments, the size of the multidimensional physical representation with DHC was smaller than that with single count header compression, logical position compression, base–offset compression and difference sequence compression. This result was true even when we included those parts of DHC not stored on the disk, but recalculated every time the DHC header was pre-loaded into the memory.

It often turns out that caching significantly improves response times. This was also found to be the case for us when the same relation is represented physically in different ways. In order to analyze this phenomenon, we proposed a model. In this model, four constants were introduced for the retrieval time from the memory ($M_m$ and $M_t$) and from the disk ($D_m$ and $D_t$). It was necessary to have four symbols as we had to distinguish between the multidimensional representation ($M_m$ and $D_m$) and the table representation ($M_t$ and $D_t$). Based on the model, necessary and sufficient conditions were given for when one physical representation results in a lower expected retrieval time than the other. Actually, with the tested benchmark databases, we found that the expected retrieval time was smaller with a multidimensional physical representation if less than 63.2% of the table representation was cached. This was true regardless of the caching level of the multidimensional representation.

We were able to infer from the model that the complexity of the algorithm could determine the speed of retrieval when (nearly) the entire physical representation was cached into the memory. A less CPU-intensive algorithm will probably result in a faster operation.

Experiments were performed to measure the constants of the model. We found there was a big difference in values between $M_m$ and $M_t$, as well as $D_m$ and $D_t$. The difference of the first two constants can be accounted for by the different CPU-intensity of the algorithms. The reason why $D_m \ll D_t$ is that the multidimensional representation requires much less I/O operations than the table representation when one cell/row is retrieved. This latter observation is in line with the *dominance of the I/O cost* rule. However, instead of counting the number of I/O operations, we chose to determine the values of $D_m$ and $D_t$ from empirical data.

We verified the model with additional experiments and found that the model fitted the experimental results quite well. There was only one slight difference with the table representation of the APB-1 benchmark database.

Finally, over the tested range of available memory, the multidimensional representation was always much faster than the table representation in terms of average retrieval time. We obtained speed up factors of up to 5 or more in the APB-1 benchmark database and up to 52 in the TPC-D database.

Based on the above results, we think, like Westmann et al. [22], that today's database systems should be extended with compression capabilities to improve their overall performance.

# Acknowledgments

I would like to thank Prof. Dr. János Csirik for his continuous support and very useful suggestions.

# Appendix

Table 6 shows the hardware and software which were used for testing. The speed of the processor, the memory and the hard disk all influence the experimental results quite significantly, just like the memory size. In the computer industry, all of these parameters have increased quickly over time. But an increase in the hard disk speed has been somewhat slower. Hence, it is expected that the results presented will remain valid despite the continuing improvement in computer technology.

Table 6: Hardware and software used for testing

| | |
|---|---|
| Processor | Intel Pentium 4 with HT technology, 2.6 GHz, 800 MHz FSB, 512 KB cache |
| Memory | 512 MB, DDR 400 MHz |
| Hard disk | Seagate Barracuda, 80 GB, 7200 RPM, 2 MB cache |
| Filesystem | ReiserFS format 3.6 with standard journal |
| Page size of B-tree | 4 KB |
| Operating system | SuSE Linux 9.0 (i586) |
| Kernel version | 2.4.21-99-smp4G |
| Compiler | gcc (GCC) 3.3.1 (SuSE Linux) |
| Programming language | C |
| Free | procps version 3.1.11 |

# References

[1] CHEN, Z. – GEHRKE, J. – KORN, F., Query Optimization in Compressed Database Systems, *ACM SIGMOD Record*, May 2001.

[2] EGGERS, S. J. – OLKEN, F. – SHOSHANI, A., A Compression Technique for Large Statistical Databases, *VLDB*, 1981.

[3] GARCIA-MOLINA, H. – ULLMAN, J. D. – WIDOM, J., Database System Implementation, *Prentice Hall, Inc.*, 2000.

[4] GOLDSTEIN, J. – RAMAKRISHNAN, R. – SHAFT, U., Compressing Relations and Indexes, *ICDE*, 1998.

[5] GRAEFE, G. – SHAPIRO, L. D., Data Compression and Database Performance, *Proc. ACM/IEEE-CS Symp. on Applied Computing*, 1991.

[6] HUFFMAN, D. A., A method for the construction of minimum-redundancy codes, *Proceedings of the IRE*, 1952.

[7] International Telecommunication Union / Line Transmission of Non-telephone Signals / Video Codec for Audiovisual Services at $p \times 64$ kbits / ITU-T Recommendation H.261
http://www.itu.org

[8] KASER, O. – LEMIRE, D., Attribute Value Reordering for Efficient Hybrid OLAP, *Proceedings of the 6th ACM international workshop on Data warehousing and OLAP*, November 2003.

[9] NG, W.-K. – RAVISHANKAR, CH. V., Block-Oriented Compression Techniques for Large Statistical Databases, *Knowledge and Data Engineering*, 1995.

[10] O'CONNELL, S. J. – WINTERBOTTOM, N., Performing Joins without Decompression in a Compressed Database System, *ACM SIGMOD Record*, March 2003.

[11] OLAP Council / APB-1 OLAP Benchmark, Release II
http://www.olapcouncil.org

[12] PENDSE, N., The origins of today's OLAP products, *(c) Business Intelligence Ltd.*, 1998.
http://www.olapreport.com/origins.html

[13] RAY, G. – HARITSA, J. R. – SESHADRI, S., Database Compression: A Performance Enhancement Tool, *International Conference on Management of Data*, 1995.

[14] SHOSHANI, A., OLAP and Statistical Databases: Similarities and Differences, *PODS*, 1997.

[15] SZÉPKÚTI, I., Multidimensional or Relational? / How to Organize an Online Analytical Processing Database, *Technical Report*, 1999.

[16] SZÉPKÚTI, I., On the Scalability of Multidimensional Databases, *Periodica Polytechnica Electrical Engineering*, 44/1, 2000.

[17] SZÉPKÚTI, I., Difference Sequence Compression of Multidimensional Databases, *to appear in Periodica Polytechnica Electrical Engineering*, 2004.

[18] TANENBAUM, A. S., Computer Networks / Third Edition, *Prentice Hall, Inc.*, 1996.

[19] TOLANI, P. M. – HARITSA, J. R., XGRIND: A Query-friendly XML Compressor, *ICDE*, 2001.

[20] TPC BENCHMARK™ D (Decision Support) Standard Specification, Revision 1.3.1.
http://www.tpc.org

[21] VASSILIADIS, P. – SELLIS, T. K., A Survey of Logical Models for OLAP Databases, *SIGMOD Record 28(4): 64-69*, 1999.

[22] WESTMANN, T. – KOSSMANN, D. – HELMER, S. – MOERKOTTE, G., The Implementation and Performance of Compressed Databases, *ACM SIGMOD Record*, September 2000.

[23] WU, W. B. – RAVISHANKAR, CH. V., The Performance of Difference Coding for Sets and Relational Tables, *Journal of the ACM*, September 2003.

[24] ZHAO, Y. – DESHPANDE, P. M. – NAUGHTON, J. F., An Array-Based Algorithm for Simultaneous Multidimensional Aggregates, *Proceedings of the ACM SIGMOD*, 1997.